\documentclass{mn2e}
\input{epsf}

\def\lsim{ \lower .75ex \hbox{$\sim$} \llap{\raise .27ex \hbox{$<$}} }
\def\gsim{ \lower .75ex \hbox{$\sim$} \llap{\raise .27ex \hbox{$>$}} }

\begin{document}
 
%TITLES AND AUTHORS
\title[Measuring the acoustic horizon with the power spectrum of clusters]
{Constraints on the dark energy equation of state from the imprint of 
baryons on the power spectrum of clusters.}

\author[Angulo  et al.]{
\parbox[t]{\textwidth}{
\vspace{-1.0cm}
R.\,Angulo$^{1,2}$,
C.\,M.\,Baugh$^{2}$,
C.\,S.\,Frenk$^{2}$,
R.\,G.\,Bower$^{2}$,
A.\,Jenkins$^{2}$,
S.\,L.\,Morris$^{2}$.
}
%\vspace*{6pt} 
\\
$^{1}$Departamento de Astronom\'\i a y Astrof\'\i sica, Pontificia
Universidad Cat\'olica de Chile, V. Mackenna 4860, Santiago 22, Chile.\\
$^{2}$Institute for Computational Cosmology, Department of Physics, 
University of Durham, South Road, Durham, DH1 3LE, UK. 
%\vspace*{-0.5cm}
}
 
\maketitle
\begin{abstract}
Acoustic oscillations in the baryon-photon fluid  
leave a signature in the matter power spectrum. The overall shape 
of the spectrum and the wavelength of the oscillations depend upon the sound 
horizon scale at recombination. Using the $\Lambda$ cold dark matter 
Hubble Volume simulation, we show that the imprint of baryons is 
visible in the power spectrum of cluster-mass dark matter haloes, 
in spite of significant differences between the halo power spectrum 
and the prediction of linear perturbation theory. 
A measurement of the sound horizon scale can constrain the dark energy 
equation of state. We show that a survey of clusters at intermediate redshift 
($ z\sim1 $), like the Sunyaev-Zeldovich survey proposed by the South 
Pole Telescope or a red sequence photometric survey with VISTA, 
could potentially constrain the sound horizon scale to an accuracy of 
$\sim 2\%$, in turn fixing the ratio of the pressure of the dark energy to 
its density ($w$) to better than $\sim 10\%$. Our approach does not 
require knowledge of the cluster mass, unlike those that depend 
upon the abundance of clusters. 
\end{abstract}

\section{Introduction }

Estimates of cosmological parameters have improved
dramatically with the advent of accurate measurements of 
temperature anisotropies in the microwave background 
radiation and the clustering of local galaxies (Efstathiou
et al. 2002; Percival et al. 2002; Spergel et al. 2003; Abazajian et al. 2004;
Tegmark et al. 2004).  A consistent picture is emerging,
the $\Lambda$CDM model, in which the geometry of the Universe is flat
(or very nearly so), with matter accounting for less than 30\% of the
required critical density (de Bernardis et al. 2000; Peacock et
al. 2001; Cole et al. 2005). Several lines of evidence indicate that
the shortfall in energy density is made up by ``dark energy'' (Riess
et al. 1998; Perlmutter et al. 1999; Efstathiou et al. 2002; Spergel
et al 2003).

The dark energy exerts a negative pressure and is responsible for the
current acceleration of the expansion of the Universe. The equation of
state of the dark energy, the ratio of its pressure to density, is
specified by a parameter $w\equiv P/\rho$. For a vacuum
energy or cosmological constant, $w=-1$ at all times; in quintessence
models, the equation of state can vary with time, i.e. $w$ is a
function of redshift (Linder 2003). Theoretically, dark
energy is poorly understood and progress will require the design and
implementation of new cosmological tests to constrain its properties.

There has been much speculation about using the wavelength of 
acoustic oscillations imprinted on the galaxy power spectrum to constrain 
the dark energy equation of state (Blake \&
Glazebrook 2003, hereafter BG03; Linder 2003; Seo \& Eisenstein 2003). 
This scale depends on the size of the sound horizon at
recombination and is the maximum distance that a
ripple in the baryon-photon fluid can travel before the sound
speed drops precipitously after recombination, stifling any further
propagation. The acoustic horizon is a standard ruler that depends 
only on physical parameters (e.g. the physical densities in matter, 
$\Omega_m h^2$, and baryons, $\Omega_b h^2$, in units of the 
critical density; $h$ is the Hubble parameter) and is determined 
from the acoustic peaks in the cosmic microwave background 
(Spergel et al. 2003). The observed wavelength of oscillation 
depends upon the geometry of the Universe and thus on the dark 
energy equation of state.

Using the ``Millennium" N-body simulation of a $\Lambda$CDM universe, 
Springel et al. (2005) demonstrated that acoustic oscillations are 
visible in the matter power spectrum at the present day, albeit in a 
modified form. The maximum relative amplitude of the oscillatory features in
the matter spectrum is around $10\%$, much smaller than the 
acoustic peaks in the power spectrum of the microwave
background (Meiskin, White \& Peacock 1999).  The detection
of these features therefore demands precision measurements of large-scale
clustering, which is only possible with a survey covering a large
volume.
The imprint of baryons in the galaxy distribution was detected in 
the power spectrum of the 2-degree field galaxy redshift survey 
(Percival et al. 2001; Cole et al. 2005) and the correlation 
function of luminous red galaxies (Eisenstein et al. 2005). These detections 
lack the accuracy needed for a competitive estimate of $w$.

Galaxy clusters are an attractive alternative to galaxies for mapping
the large-scale structure of the Universe. Rich clusters are easier to 
detect at large distances than individual galaxies and their mean 
separation is much greater. Thus, it is easier to sample large volumes 
homogeneously with clusters than with galaxies. The low space density 
of clusters might appear at first sight to make the
power spectrum difficult to measure. However, clusters have a stronger
correlation amplitude than the overall mass distribution and this
offsets their sparsity (Kaiser 1984; for an 
illustration of how clustering depends on group size see
Padilla et al. 2004).

In this letter, we first show that the imprint of baryons on the 
matter power spectrum is indeed visible in the power spectrum of 
galaxy clusters (Section~2). We do this using the ``Hubble Volume'' 
N-body simulation of a $\Lambda$CDM universe, modelling a range of 
phenomena that alter the appearance of the oscillatory features 
in the power spectrum as clustering develops. The power spectrum 
of clusters is very different from the prediction of linear perturbation 
theory so our approach represents an improvement over previous 
studies (e.g. Hu \& Haiman 2003; Wang et al. 2004). In Section~3, 
we assess the accuracy with which the sound horizon scale can potentially 
be recovered for samples of dark matter haloes that are relevant for 
forthcoming cluster surveys.  The prospects for constraining the 
dark energy equation of state are discussed in Section~4.

\section{The power spectrum of clusters}

The power spectrum of density fluctuations in the linear regime is a
well specified prediction of the cold dark matter model (Bardeen et
al. 1986). While fluctuations around the mean density remain
small, the evolution of the spectrum can be accurately
calculated using linear perturbation theory. In this case, the shape
of the spectrum is preserved but its amplitude changes. As the
perturbations grow, various processes cause the power spectrum of
galaxies or clusters to differ from the linear theory expectation: (i)
{\it Nonlinear evolution}. Mode coupling due to 
gravitational instability influences fluctuation growth and
alters the shape of the power spectrum.  (ii) {\it Biasing between
the spatial distribution of mass and its tracers}. In general, the
distribution of galaxies or clusters does not constitute a random
sampling of the mass and the two could be related in a
complicated way (e.g. Bower et al. 1993). In the simplest model, the
difference between the clustering amplitudes of the mass
and the tracers is described by a bias parameter, which can vary
with scale and redshift (e.g. Cole et al. 1998; Narayanan, 
Berlind \& Weinberg 2000).
(iii) {\it Peculiar motions and redshift errors.}
The effect of peculiar velocities on measured redshifts is to 
introduce a redshift-space distortion. On large scales, this 
takes the form of coherent flows towards massive structures that
boost the apparent amplitude of the power spectrum (Kaiser 1987). 
On small scales, random motions inside virialized objects smear out
structures, damping the power spectrum.
Errors in the redshift determination also damp the
spectrum at intermediate and large wavenumbers but do not boost the 
spectrum at low wavenumbers.

\begin{figure}
{\epsfxsize=8.5truecm
\epsfbox[65 361 438 687]{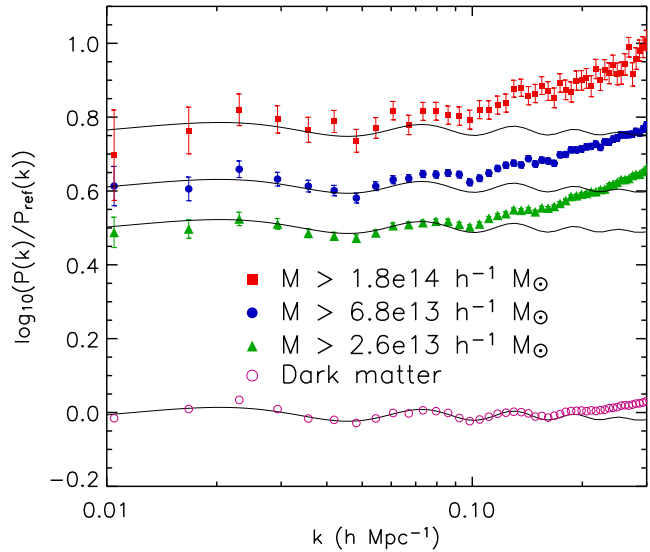}}
\caption
{
The real-space power spectrum of the dark matter (open circles) and 
various samples of dark matter haloes (filled symbols). The halo samples 
are defined by the minimum mass threshold given in the legend. 
The spectra have all been divided by a reference cold dark matter power 
spectrum without baryons. The solid lines show the linear perturbation 
theory prediction, divided by the same reference. For the halo samples, 
the lines have been multiplied by the square of an effective bias parameter.
The error bars are computed using Eq.~1.
}
\label{fig:pk}
\end{figure}

These effects are best modelled with a N-body simulation of 
hierarchical clustering. We use the $z=1$ output of the Virgo Consortium's 
``Hubble Volume'' simulation of a $\Lambda$CDM universe 
(Jenkins et al. 2001; Evrard et al. 2002).  This had 
a volume of $27h^{-3}{\rm Gpc}^{3}$ and a particle mass 
of $2.2\times10^{12}h^{-1}M_{\odot}$, making it ideally suited 
to studying the clustering of massive haloes 
(Colberg et al. 2000; Padilla \& Baugh 2002). 
Dark matter haloes are identified using the
friends-of-friends algorithm (Davis et al. 1985).

Fig.~\ref{fig:pk} shows the impact on the power spectrum
of processes (i) and (ii) above: nonlinear evolution 
and bias. The open points show the
power spectrum of the dark matter at $z=1$. 
To expand the dynamic range, the spectrum has been
divided by a cold dark matter power spectrum that does not contain 
any contribution from baryons. 
The solid line shows the linear perturbation theory 
spectrum for the simulation parameters, divided
by the same reference.  The amplitude of spectrum of 
the dark matter is systematically higher than the linear theory 
prediction for $k>0.15h{\rm Mpc}^{-1}$, a discrepancy that 
becomes more pronounced at higher wavenumbers.  The other symbols
show the spectra measured for samples of dark matter haloes
defined by different minimum mass thresholds; the amplitude of the points 
increases with the mass cut. These spectra have
been corrected for shot noise by subtracting $1/\bar{n}$, where
$\bar{n}$ is the number density of haloes. This correction
is important as the mean separation of the haloes is close to the
scale of the feature we are trying to measure. The solid lines show
the linear theory spectrum multiplied by the square of an
effective bias computed for each sample (see Padilla \& Baugh 2002). 
This model for the halo spectrum gives a good 
match to the simulation  for $k\lsim0.07 h {\rm Mpc}^{-1}$, 
but underestimates the power at higher wavenumbers.
%(The measured spectrum deviates from the linear theory prediction at
%progressively smaller wavenumbers as the halo mass defining the sample
%increases, an effect which is due, in part, to the declining space density
%of the haloes and the corresponding importance of shot noise)
Although nonlinear evolution erases the acoustic oscillations at high
$k$, the first few oscillations are still clearly visible in the
cluster power spectrum.

The impact of peculiar motions and redshift errors is
illustrated in Fig.~\ref{fig:fit}.  The symbols show the real-space 
power spectra of two halo samples. 
The dotted and dashed lines show fits to the
halo power spectrum measured in redshift-space, with and without
redshift errors respectively.  (The
fitting process is discussed in \S3.)  Redshift-space
positions are calculated in the distant observer approximation by
adding the scaled $x$ component of the peculiar motion of the centre 
of mass to the $x$ coordinate of each halo. For illustration, we 
show the impact of an error in the cluster redshift given by 
a Gaussian with variance $300~{\rm km~s}^{-1}$.  
At low $k$ we find a shift in the amplitude of the spectrum 
in redshift-space relative to real-space.  The size of this shift 
is smaller for the larger mass sample, due to its larger effective bias.

For dark matter or galaxies, redshift-space distortions
generally cause a shift in the spectrum on large scales
and damping on smaller scales relative to the real-space spectrum
(e.g. Benson et al. 2000). However, for clusters, the overall
distortion in the power spectrum is more complicated, as noted by
Padilla \& Baugh (2002).  Cluster-size haloes are in a rather different
clustering regime from the dark matter or galaxies for three 
reasons. Firstly, the mean halo separation exceeds the 
correlation length by a factor of $2$-$4$, depending upon mass. 
Secondly, haloes correspond to high peaks in the density field. 
Finally, haloes are not located in virialized structures; 
otherwise the group finder would have simply identified the supercluster 
as a dark matter halo of larger mass. 
Redshift errors depress the power spectrum at high
$k$ because they smear out structures on scales comparable
to the size of the equivalent spatial error. This effect partially
compensates for the redshift-space distortion, bringing the result
closer to the real-space estimate. The discrepancies between
the spectrum that we recover for massive haloes in redshift-space and
the naive expectations based on extrapolations of linear theory or 
results for galaxies are a good illustration of the need 
to use a full N-body calculation in order to model the power spectrum
correctly.

\section{Measuring the sound horizon}

\begin{figure}
{\epsfxsize=8.5truecm
\epsfbox[50 350 440 687]{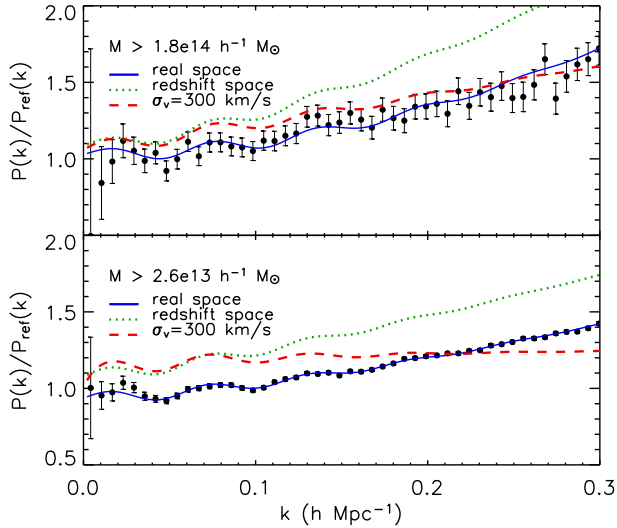}}
\caption
{
Fits to the power spectra of haloes. The points show the measured real-space spectrum and the lines show the best fits for three  
cases: real-space positions, redshift-space positions and redshift-space 
plus redshift measurement errors, as indicated by the legend. 
The panels correspond to different mass limits.
}
\label{fig:fit}
\end{figure}

\begin{figure}
{\epsfxsize=8.9truecm
\epsfbox[60 350 440 690]{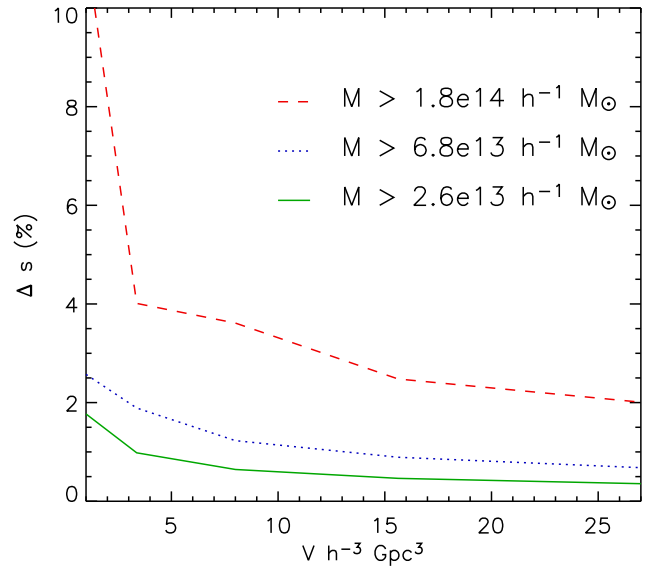}}
\caption
{The percentage error in the recovered sound horizon as a function of 
survey volume. The lines correspond to samples with different minimum 
mass thresholds, as given in the legend.}
\label{fig:err}
\end{figure}

\begin{figure}
{\epsfxsize=8.9truecm
\epsfbox[70 350 440 700]{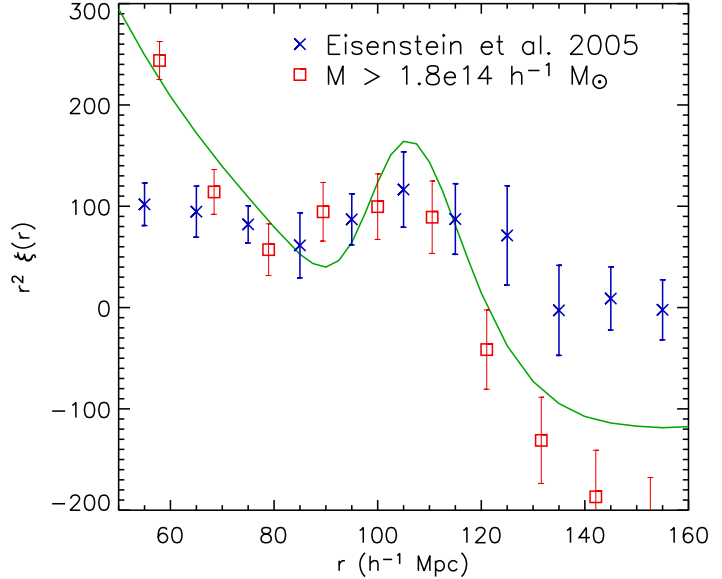}}
\caption
{The correlation function, $\xi(r)$, 
of haloes with $M > 1.8 \times 10^{14} h^{-1}M_{\odot}$ (squares) 
compared with linear perturbation theory 
(solid line), which includes the effective bias of the haloes 
and redshift-space distortions. We plot $r^{2}\xi(r)$ 
to expand the dynamic range. The correlation 
function of luminous red galaxies is also plotted for comparison (crosses;
Eisenstein et al. 2005). }
\label{fig:xi}
\end{figure}

We take a somewhat different approach to determine the sound horizon
scale to that of BG03, who fitted a parametric form to the ratio of 
the measured power spectrum to a reference spectrum without oscillatory
features. Instead, we fit to the power spectrum directly, 
using the analytic matter power spectrum presented by 
Eisenstein \& Hu (1999), which describes the linear theory 
fluctuations in the cold dark matter and baryons. We model 
the effects of nonlinear evolution using a simple approximation, 
which effectively linearizes the power spectrum by removing the nonlinear
distortion. We first divide both the measured spectrum and the 
corresponding linear theory spectrum by a smooth reference. 
The ratios obtained for the measured spectrum and the linear theory 
spectrum differ beyond a particular wavenumber, $k_{\rm nl}$ ( 
Fig.~1). We fit a linear relation to the measured power 
spectrum ratio beyond $k_{\rm nl}$, $P/P_{\rm ref} = A + B k$. 
We then divide the measured ratio by $A + Bk$ and multiply the result 
by the reference spectrum to get a linearized power spectrum. 
The free parameters in the linearization step are $k_{\rm nl}$, $A$ 
and $B$.  We then fit the linearized spectrum to the 
Eisenstein \& Hu formula. The free parameters are the sound horizon 
scale, $\Omega_m$, $\Omega_b$, $h$ and the amplitude of the
spectrum. The values of $\Omega_m$, $\Omega_b$ and $h$ are set to
those used in the simulation.

Our approach has a number of clear advantages over that of 
BG03. Firstly, BG03 fit to the measured spectrum divided by a 
reference spectrum. There is no statistical gain to be derived from this; 
we present our results as a ratio in Figs.~1 and~2 simply to 
improve the contrast of the oscillations. As a matter of fact, 
the choice of reference can compromise the fit as different 
choices for the smooth spectrum can alter the visibility of the first 
peak, leading to possible systematic errors in the derived sound 
horizon scale. Secondly, the form advocated by BG03 is an approximation 
based on a Taylor expansion of the combined transfer functions for cold dark 
matter and baryons. The parameter that BG03 equate to the sound horizon 
is actually only equivalent to this scale under certain conditions. 
Thirdly, BG03's parametric form is only applicable when
fluctuations are in the linear regime. In hierarchical models,
nonlinearities lead to deviations from the linear theory spectrum on
surprisingly large scales (see Fig.~1 and Figure ~4 of Baugh \& Efstathiou
1994). Finally, by fitting to the power spectrum rather than a ratio,
we are using information about the overall shape of the power
spectrum, including the break, which is also sensitive to the sound
horizon scale. We find that we recover a more accurate estimate of the
sound horizon scale when we fit to the power spectrum directly instead
of to a ratio. 
%These points will be dealt with in more detail in
%Angulo et al. (2005, in preparation).

The error on the spectrum depends upon the volume 
and the amplitude of the spectrum relative to the shot
noise due to the discreteness of the tracers of the
mass distribution, 
and is given by (Feldman, Kaiser \& Peacock 1994):
\begin{equation} 
\frac{\sigma}{P}  
= 2\pi 
\sqrt{\frac{1}{Vk^{2}\Delta k}} \left( \frac{1 + \bar{n}P}{\bar{n}P}
\right),
\label{eq:error}
\end{equation}   
where $V$ is the survey volume, $\Delta k$ is the width of the
bins used to average the spectrum and $\bar{n}$
is the mean number density of clusters. All quantities are comoving. The
product $\bar{n}P$ is the signal-to-noise ratio of the measured power
and depends upon $k$. We have tested Eq.~\ref{eq:error} 
against the dispersion between the 
spectra measured on subdividing the Hubble Volume simulation. 
%We have experimented with the
%maximum wavenumber over which our fitting procedure is applied. The
%recovered value of the sound horizon and the associated error are
%insensitive to the maximum wavenumber adopted over the range
%$k=0.15-0.3h{\rm Mpc}^{-1}$.

The fits to the power spectra measured for two samples of haloes
in the full simulation are shown in Fig.~\ref{fig:fit}. 
As explained above, we show the results after dividing by a smooth 
reference simply to expand the range on the y-axis. We show 
the best fit in three cases: the cluster power spectrum measured in 
real-space, in redshift-space and in redshift-space with a redshift 
error of $\sigma_{V} = 300 {\rm kms}^{-1}$. Even though the shapes of 
the spectra are quite different, our method recovers the 
theoretical value of the sound horizon scale accurately. Results for 
three samples of haloes drawn from the full Hubble Volume simulation 
output at $z=1$ are given in Table~1.
The accuracy with which we predict that the sound
horizon scale can be extracted from the halo power spectrum for
different cubical volumes is given in Fig.~\ref{fig:err}.

The sound horizon can also be measured from the correlation
function. In this case, the signature is a spike at a
comoving pair separation approximately equal to the sound horizon
scale. Fig.~\ref{fig:xi} shows the correlation function of haloes with
a minimum mass of $1.8 \times 10^{14}h^{-1}M_{\odot}$. There is a
clear peak at $\sim 105 h^{-1}$Mpc. Significant differences remain 
between the simulations results and linear perturbation theory, 
even after the effective bias of the halo sample and redshift-space 
distortions (Kaiser 1987) are taken into account.
Fig.\ref{fig:xi} also shows the  
correlation function of luminous red galaxies at $z \sim 0.35$ 
(Eisenstein et al. 2005).  The correlation amplitude and errors are
comparable to those of our sample of clusters at $z=1$, which is
purely a coincidence. The peak is better defined in the cluster
sample, further illustrating the ability of this method to yield 
information on the dark energy equation of state at different
redshifts.

\begin{table}
\caption{
{\small The accuracy with which the sound horizon can be measured.
Column (1):  The lower mass limit of the sample. 
(2) The number of haloes above this mass in the $\Lambda$CDM Hubble 
Volume output at $z=1$. (3) The accuracy (\%) with which the sound
horizon is measured from the halo power spectrum. (4) The corresponding 
accuracy with which $w$ is constrained.}  }
\begin{tabular}{cccc}
\hline
halo mass ($h^{-1}M_{\odot}$)    &  number of haloes    & $\Delta s$ (\%) & $\Delta w$ (\%)  \\
\hline
 $> 1.8 \times 10^{14}$ & \hphantom{00}45 000 &  2.1    &    9\hphantom{00}  \\
 $> 6.8 \times 10^{13}$ & \hphantom{0}380 000 &  0.7    &    3\hphantom{00}   \\
 $> 2.6 \times 10^{13}$ & 2000 000           &  0.4    &   1.5  \\
\hline
\end{tabular}\label{tab:results}
\end{table}

\section{Discussion and conclusions} 

The dark energy equation of state can be constrained by comparing 
measurements of the acoustic horizon derived from the power 
spectrum of galaxy clusters and from temperature anisotropies in the 
microwave background. 
WMAP data constrain the sound horizon scale to an accuracy of
$\sim 3\%$ (Spergel et al. 2003); this will undoubtedly improve
in the future. 
In principle, $w$ can be estimated from both the radial and 
transverse parts of the power spectrum. 
The acoustic horizon is a comoving standard ruler whereas 
cluster surveys measure redshifts and angular positions. 
Transforming redshifts to comoving distances requires the Hubble 
parameter, $H(z)$, while transforming angular separations involves the angular-diameter distance, $D_A(z)$. 
Both $H(z)$ and $D_A(z)$ depend on $w$ as well as other cosmological 
parameters. Thus, the radial and tangential components of 
the spectrum provide independent routes to $w$. 
In practice, it is unlikely that realistic surveys will have 
sufficient spectral modes for such a decomposition, so one 
may have to resort to the spherically averaged power spectrum 
used here.

The cluster power spectrum gives a purely geometrical test for $w$
based on absolute distances. By contrast, the test based 
on distant supernovae (also geometrical) relies on relative 
distance measurements, while the test based on counting clusters 
as a function of redshift depends on a combination of geometry 
and the fluctuation growth rate. 
These tests are therefore complementary and, in principle, measure 
different things.  
%Moreover, the cluster power spectrum method does not suffer 
%from some of the drawbacks of using cluster number 
%counts (Majumdar \& Mohr 2003; Wang et al. 2004). 
The utility of the counts is limited by the accuracy with 
which an observation, e.g. the Sunyaev-Zeldovich (SZ) decrement, 
can be translated into an estimate of halo mass which is 
required to link with theory 
(Majumdar \& Mohr 2003; Wang et al. 2004).
%The SZ effect depends upon the pressure of hot gas integrated over a
%column through the cluster and so is sensitive to the thermodynamic
%history of the cluster; this in turn depends upon the star formation
%histories of the cluster galaxies and on the energy they have injected
%into the ICM (Majumdar \& Mohr 2002). 
Whilst the sensitivity to cluster mass can be reduced by combining the
number counts with a measurement of the amplitude of the
cluster power spectrum, this ``self-calibration'' is still model
dependent (Majumdar \& Mohr 2004). Our method does not require a 
detailed knowledge of cluster masses as these affect the amplitude 
but not the shape of the power spectrum.
%However, the estimated accuracy of the measurement of the oscillation
%wavelength does depend upon the typical mass of the cluster sample.

All methods measure a weighted integral of $w$ over redshift
(Deep Saini, Padmanabhan \& Bridle 2003). Recent estimates at low redshift
already constrain $w$ to $\approx 20\%$ in the simplest, $w=-1$, model 
(e.g. Riess et al. 2004). Measurements at $z\gsim 1$ are therefore of
great interest. With acoustic oscillations, there is the added 
incentive that the measurement becomes easier at higher redshift
because fewer high $k$ oscillations have been erased by nonlinear 
gravitational evolution. This gain is partly offset by the greater 
difficulty in finding clusters at high redshift (though the SZ effect 
is independent of redshift). For baryon oscillations, $z\sim 1$ is a 
particularly interesting epoch because the dependence of the method on the
parameter $\Omega_m h^2$ cancels out to first order for 
$w \simeq -1$ (e.g. BG03); hence the focus on $z\sim 1$.  
We have shown that the impact of baryons on the form of the 
power spectrum of cluster-mass haloes at $z\sim1$ is 
clearly visible. For the idealised case of a cubical volume, 
it is possible, with a big enough survey, 
to measure the acoustic horizon to an accuracy of $ \approx 2\%$. 
In practice, several effects such as uncertainties in the 
cosmological parameters and the impact of the survey geometry 
on the measurement of the power spectrum will need to be taken into account. 
Our method allows a competitive constraint, $\Delta w \approx 10\%$, on 
the dark energy equation of state at $z\sim~1$ (Angulo et al, in prep).

Suitable samples of clusters at $z \sim 1$ could be
obtained in a variety of ways. First, a catalogue of clusters
on the sky is required. This could be constructed from
optical and near infrared photometry, using the red sequence of 
early type galaxies (Gladders \& Yee 2005). The photometry
could also be exploited to estimate the cluster redshift (e.g. Kodama, 
Bell \& Bower 1999). Gladders \& Yee (2005) report an accuracy of 
$\sigma_{z} \approx 0.02-0.03$ using two bands that
straddle the $4000\AA$ break at $z \sim 1$.  Large area
photometric surveys are currently being proposed. For example,
equipped with an optical camera, VISTA (http://www.vista.ac.uk) 
would be able to survey 10,000 square degrees in UVRI to 
$B \approx 25$ in 200 photometric nights and the 
Dark Energy Survey (DES, http://cosmology.astro.uiuc.edu/DES/) 
plans to cover 4,000 square degrees to a similar depth. 
Alternatively, the SZ effect could be used to identify clusters 
on the sky (Carlstrom, Holder \& Reese 2002). 
The South Pole Telescope (SPT) aims to detect 40,000 SZ clusters over 
4,000 square degrees at $z\sim 0.5-1$ (Ruhl et al. 2004). 
Photometric redshifts for these could be obtained with the DES 
which plans to cover the same area.  The objects in our most massive halo 
sample are comparable to those that the SPT will detect at $z \sim 1$.

Our test requires cluster redshifts. Obtaining spectroscopy  
for thousands of clusters is not trivial. With AA$\Omega$, 
the new multi-object spectrograph at the Anglo Australian Observatory, 
a redshift for a bright early type galaxy at $z \sim 1$ can be obtained 
in 100-200 minutes. There are typically 30 clusters per 
field of view, so a survey of 40,000 clusters would need 
$\approx 200$ clear nights; an equivalent spectrograph on an 8m
telescope would reduce this to 50 nights.
Photometric redshifts are an alternative but the reduced accuracy leads 
to a serious loss of information (Blake \& Bridle 2005). 
The error in the cluster redshift causes a damping of the power 
at $k_{x}>1/\sigma_{x}$, where $\sigma_{x}$ is the scaled error in the 
comoving cluster position. For the photometric redshift accuracy reported 
by Gladders \& Yee (2005), $\sigma_{x} \approx 140 h^{-1}$Mpc at $z=1$. 
The damping of the spectrum can be circumvented if modes 
with $k_{x} > 1/\sigma_{x}$ are discarded, thus effectively using  
the full spectrum at $k_{x}<1/\sigma_{x}$ and the transverse spectrum 
for $k_{x}>1/\sigma$. The associated reduction in the number of $k$-modes 
(by a factor of $\sim 20$; see Blake \& Bridle 2005) could 
be compensated for in several ways: (i) Increasing 
the solid angle covered. (ii) Improving the photometric 
redshift estimate by using more filters and deeper photometry. (iii) 
Using a lower mass to define the halo sample. 
Although further exploration of this idea using a mock catalogue 
of a specific survey is needed, our idealized calculation suggests 
that this is a powerful method for obtaining interesting 
constraints on the properties of the dark energy.

\subsection*{ACKNOWLEDGEMENTS}
{\small
This work was supported by the EC's ALFA scheme via 
funding the Latin American European Network for Astrophysics and 
Cosmology and by PPARC; CMB is supported by the Royal Society. 
We acknowledge discussions with Chris Blake, Sarah Bridle, 
Francisco Castander, Enrique Gazta\~{n}aga and Mike Gladders. 
}

\end{document}